\documentclass{elsart}
\usepackage{epsfig}
\begin{document}
\newcommand{\x}{\em \bf }

%\doublelinespacing
%\setlength{\baselinestretch}{1.7}
%\renewcommand{\baselinestretch}{2}

%0123456789012345678901234567890123456789012345678901234567890123456789
\mbox{}
\begin{frontmatter}
\renewcommand{\baselinestretch}{1}
\title{A linear radiofrequency ion trap
for accumulation, bunching, and emittance improvement
of radioactive ion beams}
\author[GSI]{F. Herfurth\thanksref{correspond}},
\author[GSI]{J. Dilling},
\author[CERN,montreal]{A. Kellerbauer},
\author[LMU]{G. Bollen},
\author[orsay]{S. Henry},
\author[GSI]{H.-J. Kluge},
\author[GSI]{E. Lamour},
\author[orsay]{D. Lunney},
\author[montreal]{R.~B.~Moore},
\author[GSI]{C. Scheidenberger},
\author[CERN,GSI]{S. Schwarz},
\author[GSI]{G. Sikler},
\author[JYFL]{J. Szerypo}
\address[GSI]{GSI, D-64291 Darmstadt, Germany}
\address[CERN]{CERN, CH-1211 Geneva 23, Switzerland}
\address[LMU]{Sekt. Phys., Ludwig-Maximilians-Univ.
M\"unchen, D-85748 Garching, Germany}
\address[orsay]{CSNSM-IN2P3-CNRS, F-91405 Orsay-Campus, France}
\address[montreal]{Dept. of Phys., McGill University,
Montr\'{e}al (Qu\'{e}bec) H3A~2T8, Canada}
\address[JYFL]{Dept. of Phys., Univ. of Jyv\"askyl\"a, PB 35(Y5),
FIN-40351 Jyv\"askyl\"a, Finland}

\thanks[correspond]{corresponding author: Frank.Herfurth@cern.ch,
phone/fax: +41 22 767 2780/8990}
\begin{abstract}
An ion beam cooler and buncher has been developed for the
manipulation of radioactive ion beams. The gas-filled linear
radiofrequency ion trap system is installed at the Penning trap
mass spectrometer ISOLTRAP at ISOLDE/CERN. Its purpose is to
accumulate the 60-keV continuous ISOLDE ion beam with high
efficiency and to convert it into low-energy low-emittance ion
pulses. The efficiency was found to exceed 10\,\% in agreement
with simulations. A more than 10-fold reduction of the ISOLDE beam
emittance can be achieved. The system has been used successfully
for first on-line experiments. Its principle, setup and
performance will be discussed.
\end{abstract}
PACS number: 21.10.Dr, 2.10.Bi, 07.75.+h
\begin{keyword}
Ion guide. Ion trap. Ion cooling. Ion buncher. On-line mass
spectrometry. Radioactive ion beams.
\end{keyword}
\end{frontmatter}

\section{Introduction}

The development of new techniques for the manipulation of
radioactive ion beams is actively pursued by several groups
worldwide. One of the main objectives is a better matching of the
properties of the radioactive ion beams to the specific
requirements of the experiments. Here ion trap techniques have
started to play an increasingly important role, in particular for
the accumulation, cooling, and bunching of these beams. Both,
Penning traps \cite{Boll96a} and radiofrequency multipole ion
traps or guides \cite{Moore92} can fulfill this task. In addition,
Penning traps offer high-resolution mass separation and can be
used for beam purification~\cite{HRH97}. Radiofrequency multipole
ion guides have been employed in ion chemistry and molecular
physics for many years~\cite{Teloy74,Gerlich92}. Now they have
gained increasing importance in the field of nuclear physics,
where they are used for guiding charged nuclear reaction products
from high-pressure gas cells into high-vacuum regions in order to
form low-energy radioactive ion beams
\cite{LISOL,Wada,JYFLguide,ANL} or for enabling ultra-sensitive
laser experiments on radioactive ions \cite{Backe97}.

A rather new application is the use of radiofrequency multipole
systems for the manipulation and improvement of radioactive ion
beams as they are  available from on-line mass separators. At the
ISOLDE facility~\cite{PSB-ISOLDE} at CERN, a beam accumulator,
cooler, buncher, and emittance improver based on a linear
radiofrequency quadrupole (RFQ) ion trap has been realized and
used for on-line physics experiments. The system is installed at
the Penning trap mass spectrometer ISOLTRAP \cite{boll96b}. Its
task is the conversion of the continuous 60-keV ISOLDE ion beam
into low-energy, low-emittance ion bunches that can be transferred
with high efficiency into the mass spectrometer.

\section{The ISOLTRAP facility}
\begin{figure}
{\hspace{.2cm}
 \begin{center}
 \epsfig{file=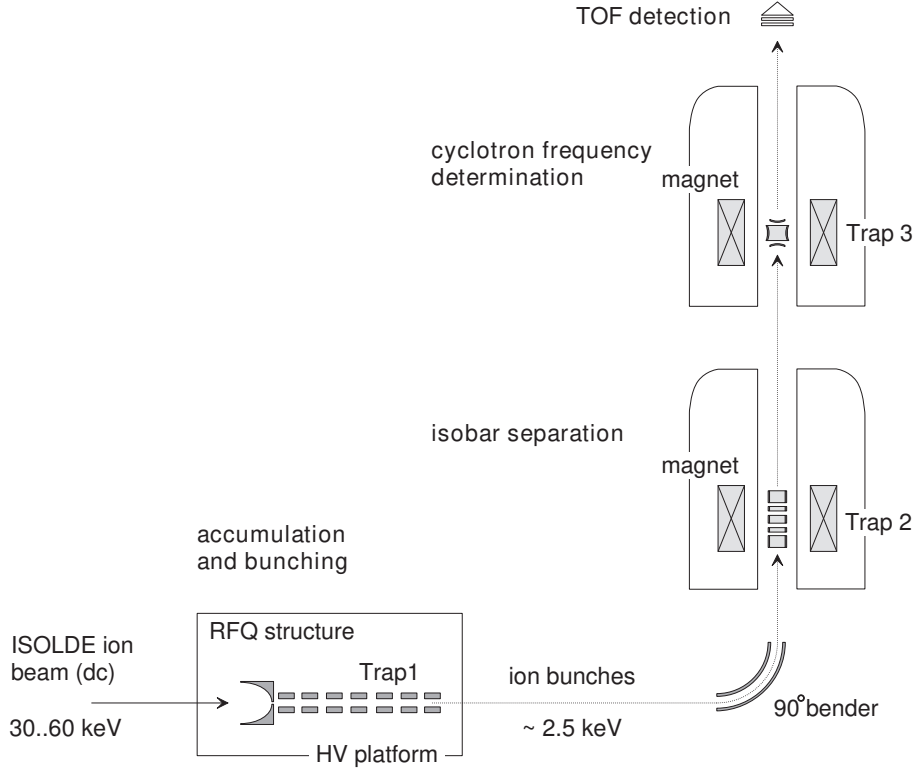,width=12cm}
 \caption{\label{f_isoltrap}
 Experimental setup of the ISOLTRAP Penning trap mass spectrometer
 and the linear RFQ ion beam buncher.}
 \end{center}
}
\end{figure}

Figure \ref{f_isoltrap} shows an overview of the layout of the
ISOLTRAP Penning trap mass spectrometer together with the
radiofrequency quadrupole ion beam buncher. The ISOLTRAP
spectrometer itself consists of two Penning traps and has been
described in detail in earlier publications \cite{boll96b,HRH97}. Here
only a short description of the tasks of the Penning traps and of the
basic operation will be given.

Low-energy ion pulses delivered by the RFQ ion beam buncher are
captured in the first Penning trap.
The main task of this trap is the purification of the ion bunch from
contaminating ions. A mass selective buffer gas
cooling technique \cite{sava91,koen95} is employed, which is based on
the simultaneous application of a buffer gas and an RF excitation of
the motion of the stored ions. This technique allows operation of the
trap as an isobar separator with a resolving power of up to
$R\approx 10^5$ for ions with mass number $A\approx100$ \cite{HRH97}.

The second trap is a high-precision trap \cite{boll96b,beck90}
used for the mass measurement of the ions delivered by the first
Penning trap. The mass measurement is carried out via a
determination of the cyclotron frequency $\omega_{\mathrm{c}} =
\frac{q}{m} B$ of an ion with mass $m$ and  charge $q$ in a
magnetic field of known strength $B$. Using this technique, to
date more than 150 isotopes have been investigated with a mass
accuracy of $\delta m/m \approx 1\cdot 10^{-7}$ (see for example
\cite{Beck97,Ames99}).

For the ISOLTRAP experiment (and any other ion trap experiment
connected to an on-line mass separator), the most critical issue
is the efficient transfer of the radioactive ion beam, which is
typically of a few tens of keV in energy, into the ion trap. In an
earlier version of ISOLTRAP, a stopping--reionization scheme was
used in order to obtain a low-energy beam of radioactive ions.
This approach limited the applicability of the Penning trap mass
spectrometry to those elements that can be efficiently
surface-ionized. Already several years ago, a development of an
ion beam accumulator based on a Paul trap system was started. Such
a system \cite{Moore92,SchwarzPhD} was indeed realized and has
been used successfully at the ISOLTRAP experiment
\cite{SchwarzPhD}. However, it was found that the acceptance of
the Paul trap system was not well matched to the ISOLDE beam
emittance and that the ejection of the ions out of the system was
very critical. The linear ion trap system to be discussed in this
paper follows the same concept as the earlier Paul trap system,
but avoids these difficulties.

\section{Principle of the ISOLTRAP ion beam cooler and buncher}

\begin{figure}
{\hspace{.2cm}
 \begin{center}
 \epsfig{file=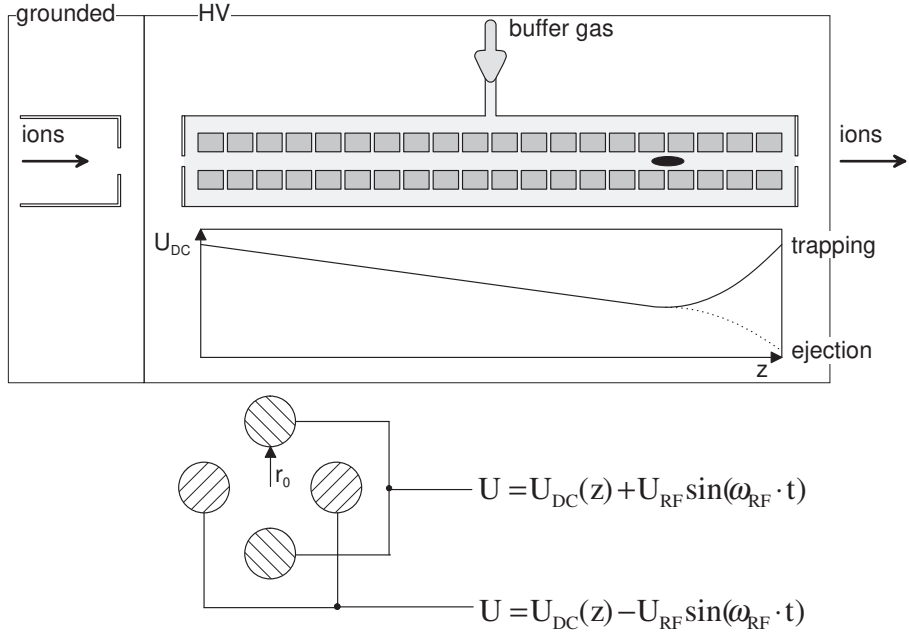,width=12cm}
 \caption{\label{f_buncher_principle}
 Basic scheme of a radiofrequency ion beam cooler and buncher. The
 upper figure shows a schematic side view of such a system together
 with the potential
 along the symmetry axis. The lower figure shows the four rods and
 their supply with radiofrequency and DC voltages.}
\end{center}
}
\end{figure}

The principle of the ISOLTRAP ion beam cooler and buncher is
illustrated in Fig. \ref{f_buncher_principle}. The 60-keV ISOLDE
ion beam is electrostatically retarded to an energy of a few eV
and injected into a linear radiofrequency quadrupole ion trap,
which is filled with a buffer gas. The trap system consists of
four segmented rods to which radiofrequency voltages are applied
so as to obtain a transversely focusing force. The segmentation of
the rods allows the creation of a DC electric field along the axis
of the system. Ions entering the linear trap will lose transverse
and longitudinal energy due to collisions with the buffer gas.
They are finally accumulated in the potential well at the end of
the system forming an ion bunch. By switching the potential of the
last rod segments, as indicated in the figure, the ion bunch can
be extracted. For a discussion of the principle of similar devices
used as an ion guide see \cite{Lunney99,Moore2000}.

\subsection{Linear radiofrequency quadrupole ion traps}

The confinement of an ion in a linear RFQ trap as presented here
is achieved by combining a static trapping potential
$V_{\mathrm{DC}}$ along the axis of the system with a radial
confinement. The latter results from the interaction of the
trapped ion with a radiofrequency quadrupole field generated by
the RF voltages applied to the quadrupole rods of the system.

The radial confinement in an exactly quadrupolar field can be
described in terms of Mathieu equations \cite{Dawson,PaulNobel},
which also define the stability criteria of the motion.  It is
useful to consider the resulting motion as being that of a
particle in a pseudopotential well \cite{Dehmelt} of depth
$V_{\mathrm{RF}}$. While the electric field configuration of the
linear trap produces motions that cannot be exactly described in
terms of the Mathieu functions (due to the strong radial and
azimuthal dependence of the axial field), it is still useful to
consider the first-order solution of the motion as a linear
superposition of the DC axial field potential and the RF
quadrupole pseudopotential.

The pseudopotential is generated by radiofrequency voltages with
amplitudes $\pm U_{\mathrm{RF}}$ and angular frequency
$\omega_{\mathrm{RF}}$. As shown in
Fig.\,\ref{f_buncher_principle}, these voltages are applied with
180$^\circ$ phase difference to pairwise connected opposite
elements of a quadrupole rod system.  The separation between the
surfaces of opposite electrodes is $2r_0$. The resulting
pseudopotential is given by

\begin{equation}
V_{\mathrm{RF}}(r)=\frac{q\cdot U_{\mathrm{RF}}}{4 r_0^2}r^2
\end{equation}
where
\begin{equation}
q=4\frac{eU_{\mathrm{RF}}}{m r_0^2 \omega_{\mathrm{RF}}^2}
\end{equation}
is the relevant Mathieu parameter and $e$ and $m$ are the charge
and mass of the stored ion. The solution of the Mathieu equation
shows that the motion is stable as long as $q<0.908$.  In the
radiofrequency field, the ion performs a micro-motion at the
frequency $\omega_{\mathrm{RF}}$ of the field and a macro-motion
that can be understood as an oscillation in the pseudopotential
$V_{\mathrm{RF}}$. To a good approximation for $q<0.6$, its
oscillation frequency is
\begin{equation}
\omega_{\mathrm{m}}=\frac{q}{\sqrt{8}}\omega_{\mathrm{RF}}~~~.
\end{equation}

As an example, a singly charged ion with mass number $A=39$ in a
four-rod structure with $r_0=6$\,mm operated with
$\omega_{\mathrm{RF}}=2\pi\cdot1$\,MHz, $U_{\mathrm{RF}}=80$\,V
will experience a depth of the radial trapping potential of about
$ 11$\,V in which it oscillates with $\omega_{\mathrm{m}}=
196$\,kHz.

As illustrated in Fig.\,\ref{f_buncher_principle}, the axial
confinement is provided by applying DC voltages to the different
segments of the rods in order to create a potential well along the
trap axis. The minimum of the potential curve can be approximated
by a parabola $V_{\mathrm{DC}}(r=0,z)=
(U_{\mathrm{DC}}/z_0^2)\cdot z^2$ defined by the characteristic
length $z_0$ and voltage $U_{\mathrm{DC}}$. Since $\bigtriangleup
V_{\mathrm{DC}}(r,z)=0$, the corresponding axisymmetric quadrupole
potential is

\begin{equation}
V_{\mathrm{DC}}(r,z)=\frac{U_{\mathrm{DC}}}{z_0^2}\left
(z^2 - \frac{r^2}{2} \right )~~~.
\end{equation}

In the region where the DC potential along the trap axis has its
minimum, the radial confinement due to the radiofrequency field is
counteracted by the repelling radial part of the DC potential. The
overall potential in the trap minimum then becomes

\begin{equation} \label{V_average}
\bar V(r,z)=
\frac{U_{\mathrm{DC}}}{z_0^2}z^2 +
\left (\frac{q}{4} \frac{U_{\mathrm{RF}}}{r_0^2}-\frac{U_{\mathrm{DC}}}
{2 z_0^2}\right )
r^2~~~.
\end{equation}

With the parameters used above and $U_{\mathrm{DC}}/z_0^2=
10\,$V/cm$^2$, the radial trapping potential well is reduced by
about 1.8 V compared to the RF-only case.  From
Eq.\,\ref{V_average} the condition
\begin{equation} \label{U_critical}
U_{\mathrm{RF,min}}=r_0^2\cdot \omega_{\mathrm{RF}}
 \sqrt{\frac{m}{e}\frac{U_{\mathrm{DC}}}{2z_0^2}}
\end{equation}
can be derived for the minimum RF
voltage required for three-dimensional ion trapping.

\subsection{Buffer gas cooling}

After the ions have entered the linear radiofrequency quadrupole,
they interact with the buffer gas. The ions are elastically
scattered at the buffer-gas atoms transferring part of their
energy to them. Thus, the motion of the ions is damped until they
finally come into thermal equilibrium with the buffer gas in the
minimum of the trapping potential.

To study the overall cooling process, it is sufficient to describe
the action of the gas as that of a viscous force. However, for an
understanding of ion loss mechanisms and final ion temperatures, a
microscopic modelling of the deceleration and cooling process
including the radiofrequency field is required. Various
calculations of both kinds have been carried out for the system
presented here. In the following the most basic aspects of buffer
gas cooling in such a system will be discussed.

For low ion energies (less than a few eV), the damping of the ion
motion is dominated by the long-range interaction of the ion with
buffer gas atoms, polarized by this ion. This interaction results
in an average damping force
\begin{equation}
\vec{F}=-\delta \cdot m\cdot \vec{v}~~~~,
\end{equation}
where $m$ and $v$ are the mass and the velocity of the ion.
The damping coefficient
\begin{equation}
\delta =\frac{e}{m}\frac{1}{\mu}
        \frac{p/p_{\mathrm{N}}}{T/T_{\mathrm{N}}}
\end{equation}
is proportional to the gas pressure $p$ (in fractions of the
normal pressure $p_{\mathrm{N}}$) and inversely proportional to
the temperature $T$ (in fractions of the normal temperature
$T_{\mathrm{N}}$) and the reduced ion mobility $\mu$
\cite{McDaniel}, which for low kinetic energies of the ions is
constant for a given ion species and type of buffer gas. For
kinetic energies $>1$\,eV the ion mobility decreases (see
discussion in \cite{Lunney99}). For the purpose of this paper it
is sufficient to take the constant low-energy values as an upper
limit for the ion mobility.

\begin{figure}
{\hspace{.2cm}
 \begin{center}
 \epsfig{file=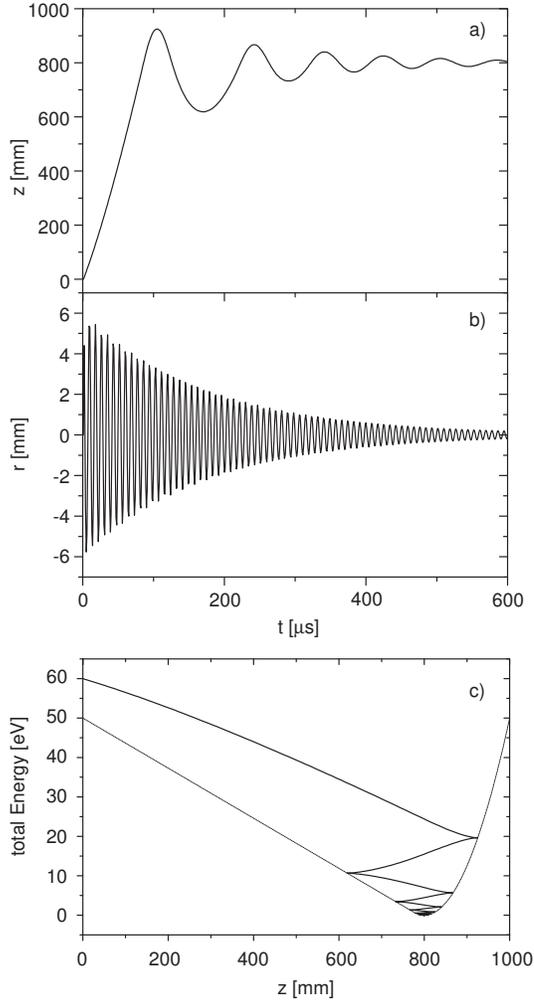,width=7cm}
 \caption{\label{f_cooling_principle}
 Simulation of the accumulation and cooling process in a linear
 RFQ trap: a) the
 axial oscillation as a function of time,
 b) the radial oscillation as a function of  time, and c)
    the total axial energy as a function of the axial position
    (shown together with the axial potential).}
 \end{center}
}
\end{figure}

Figure \ref{f_cooling_principle} illustrates the cooling and
accumulation process in a linear trap with an axial potential as
shown in Fig.\,\ref{f_buncher_principle}. The simulation has been
performed for a $^{39}$K$^+$ ion entering the system with an
initial kinetic energy of $E_{\mathrm{kin}}=10$\,eV. The axial
potential well depth is $V_{\mathrm{z0}}=50$\,V and the trap has a
total length of $L=1$\,m with the potential minimum at
$l_0=0.8$\,m. A helium buffer gas pressure of
$p_{\mathrm{He}}=10^{-2}$\,mbar has been used. With an ion
mobility $\mu ({\rm K^+ -He})=2.15\cdot 10^{-3}$\,m$^2$/Vs,  the
damping constant reaches a value of $\delta = 11500$\,s$^{-1}$. As
can be seen in Fig.\,\ref{f_cooling_principle}(a) and (b), both
the axial and the radial oscillation amplitude of the ion are
damped very quickly. Even more illustrative is
Fig.\,\ref{f_cooling_principle}(c) which shows the total energy of
the ion as a function of the axial position. Already during its
first oscillation in the system the ion loses more than 50\,\% of
its initial total energy. The equilibrium with the buffer-gas
temperature is reached within about 1\,ms.

\begin{figure}
{\hspace{.2cm}
 \begin{center}
 \epsfig{file=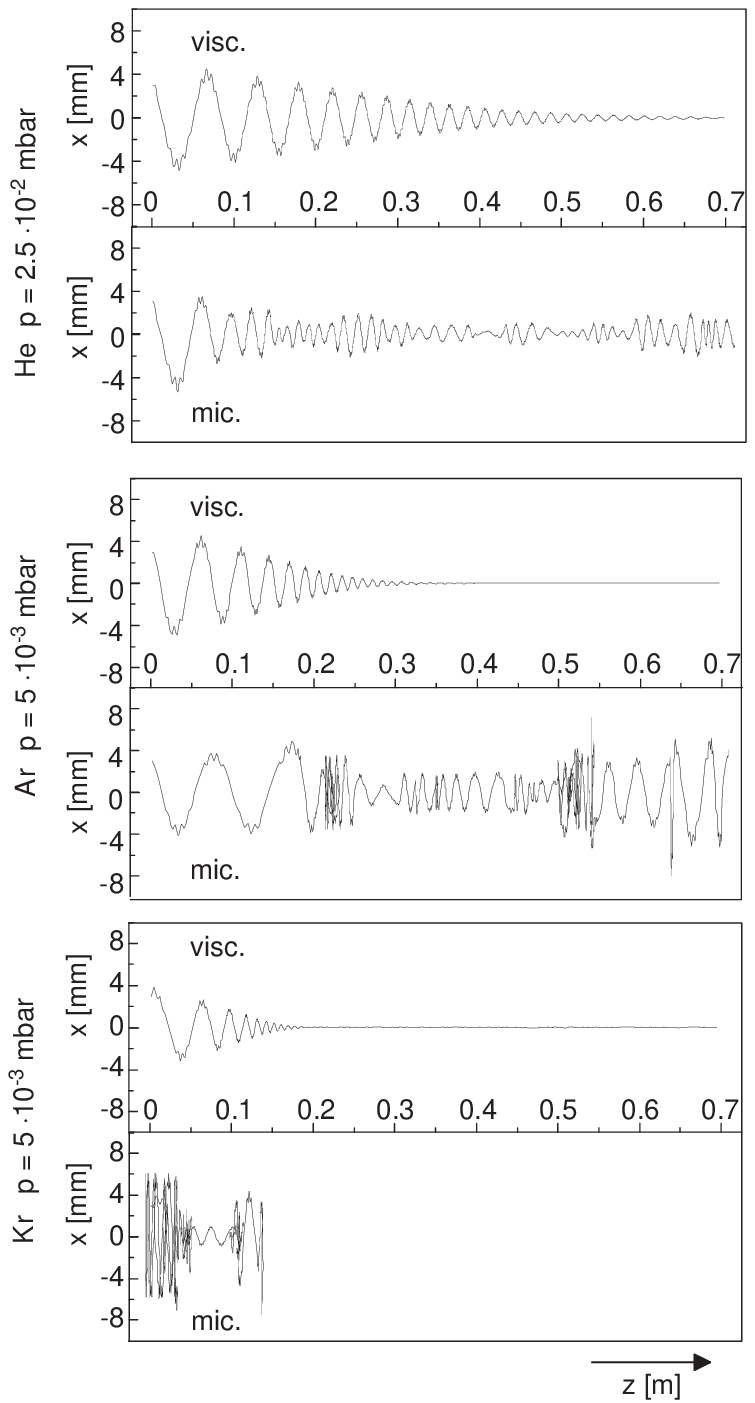,width=7cm}
 \caption{\label{f_visc_micro_comparison}
 Comparison of ion damping in a linear cooler as obtained by
 calculations based on a viscous damping approach ({\x visc}) and
 a microscopic description ({\x mic}). Shown is the radial position
 of the ions as a function of their axial position. The calculations
 have been  performed for K$^+$ in He, Ar, and
 Kr at different  pressures.}
\end{center} }
\end{figure}

The viscous damping approach presented above is valid at low
velocity for ions with masses much heavier than the mass of the
buffer gas atoms. Figure\,\ref{f_visc_micro_comparison}
illustrates the limits of this approach. A comparison of the ion
motion in the first (linear potential) part of the quadrupole
system is shown, as calculated under the assumption of viscous
damping and in a full Monte-Carlo calculation. In this calculation
the trajectories of several thousand ions have been investigated
using realistic interaction potentials for the collision of the
ions with the buffer-gas atoms (details of this microscopic
calculation will be discussed in an upcoming publication
\cite{Schwarz2000misc}). In the case of Cs ions in He, the
validity of the viscous damping description is nicely confirmed,
and even for Cs$^+$ in Ar, good agreement is obtained in the time
average. However, in the case of K$^+$ in Kr, the motion becomes
unstable very quickly and the ion is radially ejected out of the
system. The reason is that the micro-motion of an ion colliding
with a heavier atom can make a phase jump with respect to the
radiofrequency which leads to an increase of the kinetic energy.
This process is called RF-heating. In the case of light ions and a
heavy gas, it is the predominant cause of ion loss from RFQ traps.

\begin{figure}
{\hspace{.2cm}
 \begin{center}
 \epsfig{file=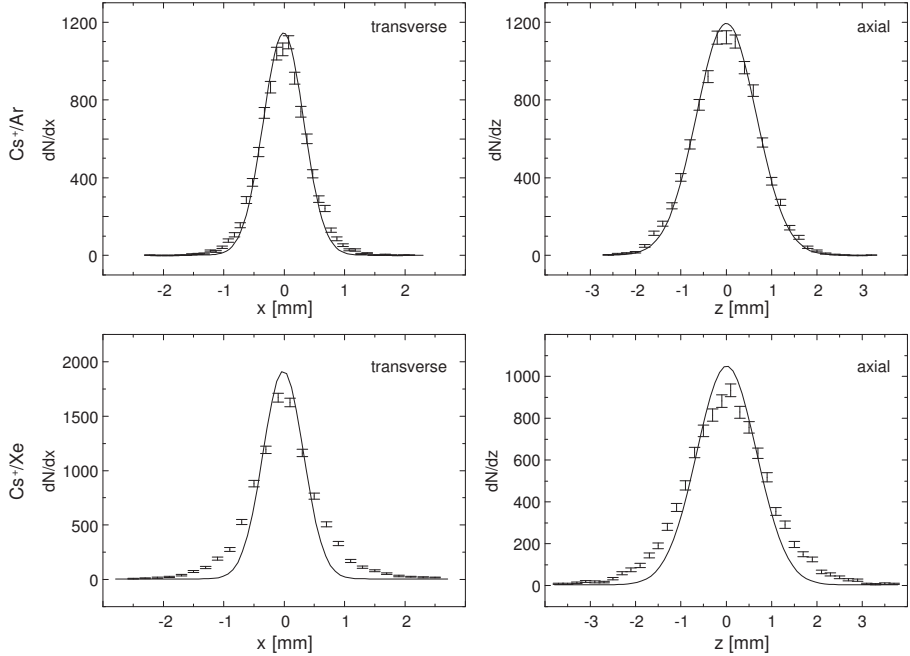,width=12cm}
 \caption{\label{f_ion_distributions}
 Transverse and axial amplitude distribution of Cs ions in the
 potential minimum of
 the linear ion trap. The solid curves show distributions for an
 ion cloud temperature of 300\,K in the total
 static (pseudo)-potential. The points are the result of a
 microscopic  calculation  taking the RF field and ion-atom collisions
 into
account.}
\end{center} }
\end{figure}

In order to determine the properties of ion bunches extracted from
the trap, the final spatial distribution of the trapped and cooled
ions is of interest. As a first approach we take the form of the
overall potential in the trap center from Eq.\,\ref{V_average} and
the parameters as used above and assume that the ions reach a
final temperature $T = 300$\,K equal to that of the buffer gas.
Assuming Boltzmann distributions, we obtain axial and transverse
distributions as shown as solid curves in
Fig.\,\ref{f_ion_distributions} for Cs$^+$ ions. The points are
the result of a microscopic calculation. It can be seen that the
oscillating electrical field and the collisions with the gas have only 
a very small effect on the ion distribution for Cs$^+$ in Ar but
a significant one for Cs$^+$ in Xe. Practically no effect 
is seen for Cs$^+$ in He, not shown in the figure here.

\subsection{Ion injection and extraction}

\subsubsection{\label{sec_decelinj}Deceleration and injection}

In order to stop an energetic ion beam in the linear ion trap by
gas collisions, first the beam has to be decelerated to low
energies (in the order of 10\,eV). This can be accomplished by
placing the whole ion trap on a potential slightly below the
corresponding ion beam energy. With reduction of energy the
divergence of the beam increases. Therefore, it depends on the
original emittance of the ion beam how far the energy of the beam
can be reduced by electrostatic means. For example, an ion beam of
$E_0 = 60$\,keV with an emittance of $\epsilon = 35 \, \pi
$\,mm\,mrad that is decelerated to $E_1=20$\,eV and focused to fit
through a 5-mm-diameter opening has a maximum divergence of
$\theta\approx 45^o$ and a maximum transverse energy of about 10
eV. Whether such a beam can be injected into a linear trap without
losses depends on the acceptance of the system. The latter is
determined by the transverse dimensions and the transverse
focusing force inside the trap.

\begin{figure}
 {\hspace{.2cm}
 \begin{center}
 \epsfig{file=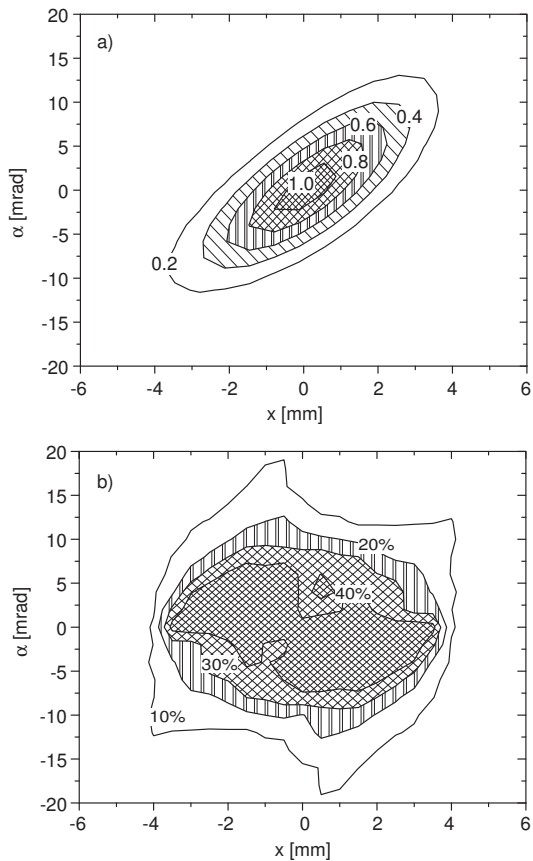,width=7cm}
 \caption{\label{f_ISOLDE_acceptance}
 a): emittance diagram of the ISOLDE ion beam. Shown is the
 ISOLDE beam intensity (in arbitrary units) as a function of beam
 displacement
 $x$ and angle $\alpha$. b): acceptance diagram of the
 ISOLTRAP ion cooler and buncher for $x$ and $\alpha$. Shown
 is the probability of successful ion injection, obtained by averaging
 over the results of the beam parameters $y$ and $\beta$ and
 all RF phases.}
 \end{center}
 }
\end{figure}

For the ion beam buncher discussed here, a number of injection
calculations have been performed in order to determine the
acceptance of the system. As an example,
Fig.\,\ref{f_ISOLDE_acceptance} shows a comparison of the
transverse acceptance of the ISOLTRAP ion beam buncher and the
transverse emittance of the ISOLDE ion beam, both at beam energies
of 30 keV. The diagrams are calculated at a position 140 mm
upstream from the 6-mm hole of the retardation electrode.

The emittance diagram was obtained from beam transport
calculations (GIOSP \cite{GIOSP}) of the ISOLDE beam from the ion
source to the apparatus. The focusing parameters of ISOLDE beam
line elements were varied until maximum overlap with the buncher
acceptance was obtained. For the acceptance diagram, a Monte-Carlo
simulation for the ion injection into the buncher was performed
(for the RFQ parameters see Table \ref{t_buncher_parameters}). An
ion was considered to be confined when it passed the first 200 mm
of the quadrupole rod system without hitting an electrode. The
calculation was repeated for various phases of the radiofrequency
field. For each $x$ and $\alpha$ value an average of the result
was taken including the beam dimensions $x$ and $\beta$. The
result is the acceptance diagram shown in
Fig.\,\ref{f_ISOLDE_acceptance}(b). From this acceptance and the
ISOLDE beam emittance a value of $\epsilon_{\mathrm{calc}}\approx
35$\,\% is obtained for the theoretical capture efficiency.

\subsubsection{Extraction and acceleration}

\begin{figure}
{\hspace{.2cm}
 \begin{center}
 \epsfig{file=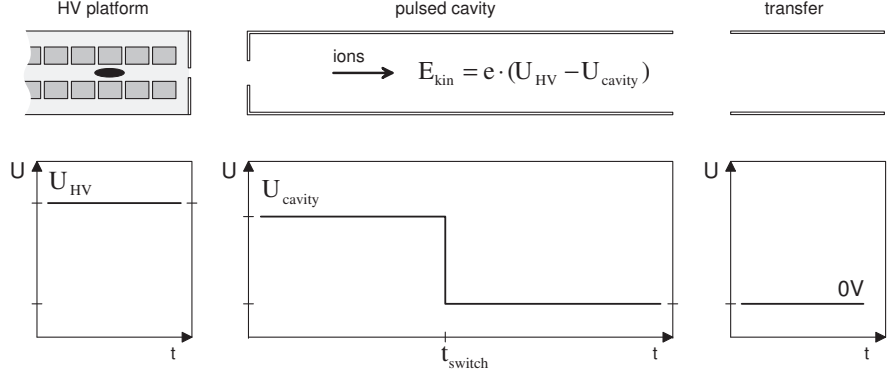,width=12cm}
 \caption{\label{f_drifttube_principle}
 Principle of creating variable energy ion pulses by employing a
 pulsed cavity}
 \end{center}
}
\end{figure}

When the ions have been accumulated in the trap potential minimum,
they can be extracted by switching the potential as indicated in
Fig.\,\ref{f_buncher_principle}. With appropriate voltages applied
to the electrodes, ion pulses not longer than a few micro seconds
can be generated. The beam properties of the extracted ion pulses
depend on the temperature and spatial distribution of the ion
cloud. The time structure depends in addition on the shape of the
potential used to extract the ions. If no further measures are
taken, the ion pulse leaving the trap at the potential of the HV
platform will be accelerated towards ground potential and once
again reach the energy of the injected ion beam. For the transfer
into the ISOLTRAP Penning traps, low-energy ($\approx 2.5\,keV$)
ion pulses are required.

An elegant way to modify the potential energy of an ion pulse is
to have the ions enter a pulsed cavity. The principle is
illustrated in Fig.\,\ref{f_drifttube_principle}. After having
left the linear ion trap at a potential $U_{\mathrm{HV}}$ the ion
pulse is accelerated towards a cavity at a potential
$U_{\mathrm{cavity}}$ to gain a kinetic energy
$E_{\mathrm{kin}}=e\cdot (U_{\mathrm{HV}}-U_{\mathrm{cavity}})$.
When the ion pulse reaches the field-free region inside the
cavity, the latter is switched to ground potential and the ions
leave it without any further change of their kinetic energy.

\section{The experimental setup}

\begin{figure}
{\hspace{.2cm}
 \begin{center}
 \epsfig{file=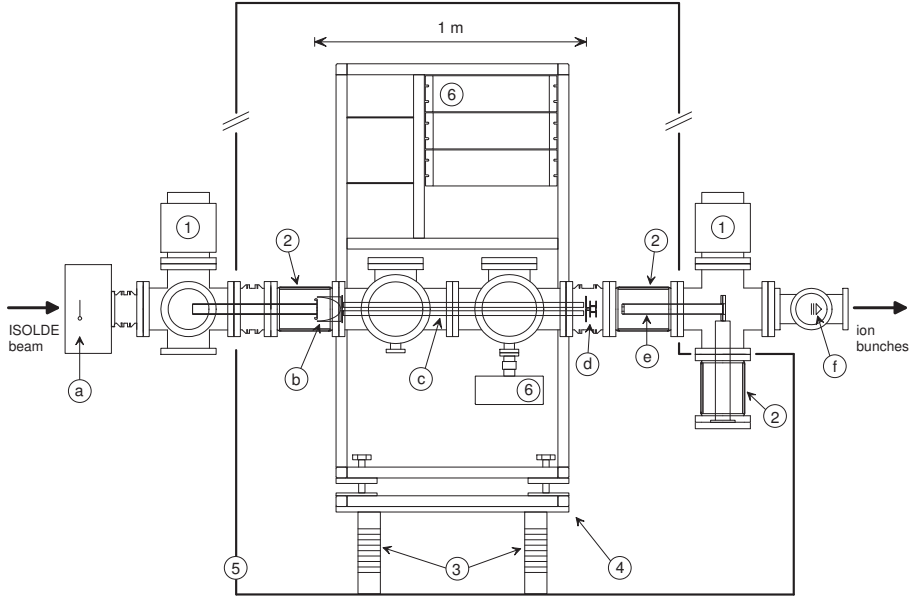,width=12cm}
 \caption{\label{f_buncher_setup}
 Experimental setup of the ISOLTRAP ion beam cooler and buncher.
 The main components are:
 (1) turbo-molecular pump, (2) insulator,
 (3) insulating support, (4) HV platform, (5) grounded cage,
 (6) DC and RF supplies,
 (a) beam scanner and Faraday cup, (b) deceleration section,
 (c) linear ion trap,
 (d) extraction section, (e) pulsed cavity, (f) MCP detector.
 }
 \end{center}
}
\end{figure}

\begin{table}[b]
\caption{\label{t_mechanics} Dimensions of the ISOLTRAP ion beam
buncher}
\begin{center}
\vspace*{1cm} \tiny \scriptsize
\begin{tabular}{lr}
\hline Element                          & Dimension [mm] \\ \hline
\hline diameter of injection hole          &        6\\ diameter
of extraction hole         &        6\\[3mm] distance between
opposite rods ($2\cdot r_0$)  &       12\\ diameter of rod
segments & 18\\ total length of quadrupole rods  &    881.5\\
length of rod segments           &         \\ \# 1, 2 & 20.5\\ \#
3 -- 19                        &     41.5\\ \# 20 & 20.5\\ \# 21
&     41.5\\ \# 22 -- 25 & 10.0\\ \# 26 &     20.5\\[3mm] length
of pulsed cavity & 380\\ \hline
\end{tabular} \tiny
\normalsize
\end{center}
\end{table}

Figure\,\ref{f_buncher_setup} shows the setup of the ISOLTRAP ion
beam cooler and buncher. On one side the system is connected to
the ISOLDE beam line system and on the other side to the ion beam
transport system of ISOLTRAP. A 60-keV test ion source and a beam
switchyard (not shown in the figure) are installed upstream in the
ISOLDE beam line in order to allow test measurements without the
ISOLDE ion beam. Beam intensities and profiles can be measured
with a needle beam scanner and a Faraday cup in front of the ion
beam buncher. Most of the relevant parts of the cooler and
buncher, including the electronics and the gas inlet system, are
placed on a 60-kV high-voltage platform in a high-voltage cage.
Ceramic insulators separate the vacuum system on the HV platform
from the beam lines on ground potential. Efficient pumping is
achieved by turbo-molecular pumps at ground potential placed close
to the insulators.

The ion-optical elements of the cooler and buncher can be grouped
into three functional sections. In the first section the
deceleration of the ISOLDE ion beam takes place, the second
section is the linear ion trap, and in the third section the
extracted ion bunches are accelerated to the desired transport
energy. All electrodes are made of stainless steel. Alumina or
glass ceramic (Macor) are used for insulating parts. The
electrical connections to the ion-optical elements are made via
UHV multi-pin feedthroughs and Kapton-insulated wires. Table
\ref{t_mechanics} summarizes some important mechanical parameters
of the system which will be discussed below in more detail.

\subsection{The deceleration and injection section}

\begin{figure}
{\hspace{.2cm}
 \begin{center}
 \epsfig{file=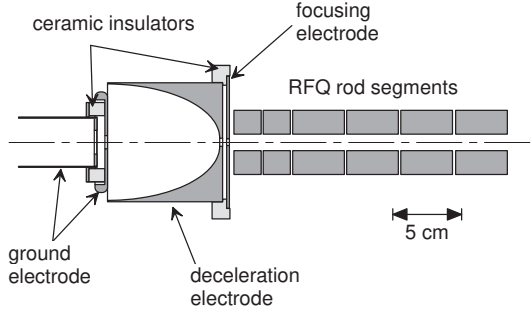,width=7cm}
 \caption{\label{f_retardation_optics}
 Ion optical elements for the deceleration of the 60 keV ion beam and
 first part of the
 linear ion trap.}
 \end{center}
}
\end{figure}

Figure\,\ref{f_retardation_optics} shows the design of the
deceleration electrode system and its connection to the linear ion
trap system discussed below. This deceleration electrode system is
the copy of a prototype described in more detail in
\cite{Moore2000}. The ISOLDE ion beam enters through the ground
electrode which is equipped with an insulated diaphragm for
current measurements. The shape of the deceleration electrode is
designed to focus the beam through the 6-mm opening at its end. An
additional electrode and the first segments of the four-rod
structure are used to obtain the final retardation and to focus
the beam towards the axis of the system.

For the presented cooler and buncher ion optical simulations were
performed so as to determine the optimal position of the ground
electrode with respect to the deceleration electrode and to get a
set of voltages that give the best injection efficiency into the
trap system. These voltages were used as start values for the
experimental optimization. A set of typical voltages for the
deceleration section are listed in
Table\,\ref{t_buncher_parameters} below together with parameters
for the other sections.

\subsection{The linear ion trap}
\begin{figure}[b]
{\hspace{2cm}
 \begin{center}
 \epsfig{file=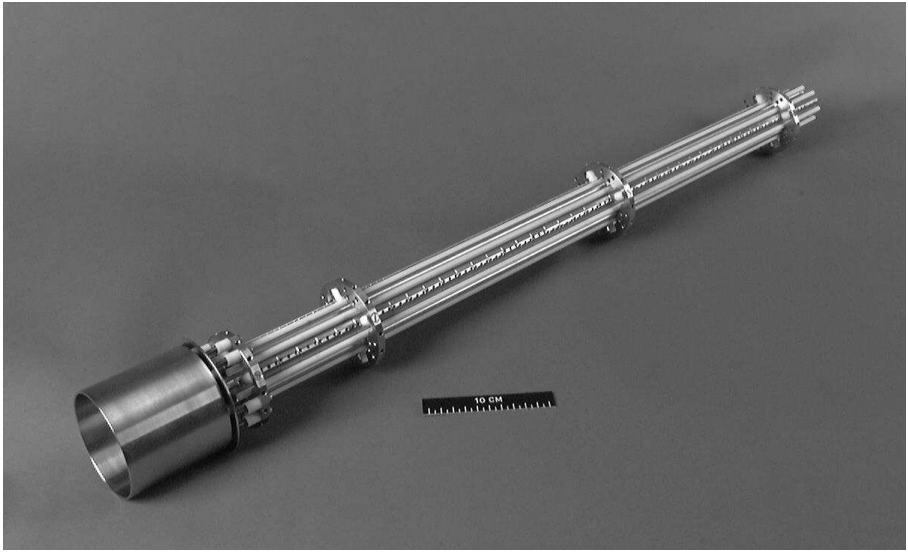,width=12cm}
 \caption{\label{f_photo_buncher}
 Photograph of the assembled four-rod electrode system of the
 ISOLTRAP ion beam
 cooler and buncher and the mounted deceleration electrode.}
 \end{center}
}
\end{figure}
\begin{figure}
{\hspace{.2cm}
 \begin{center}
 \epsfig{file=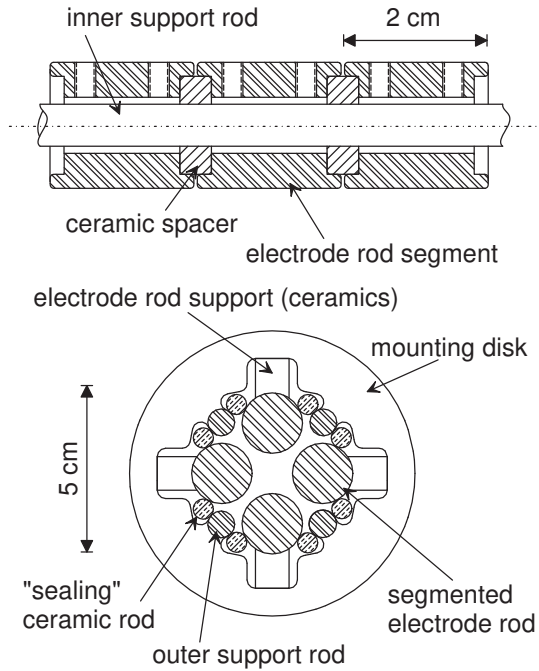,width=7cm}
 \caption{\label{f_detail_rod}
 Detail of the rod structure. The segments are insulated from each
 other by ceramic spacers. The four rods are mounted to disks
 as shown in the lower part.}
\end{center}
}
\end{figure}

Figure\,\ref{f_photo_buncher} shows a photograph of the assembled
electrode system of the linear ion trap with the deceleration
electrode mounted. Figure\,\ref{f_detail_rod} shows a detailed
view of a part of a rod and how the cylindrical electrode segments
are mounted together. The quadrupole rods have a total length of
$881.5$\,mm. Each rod consists of 26 segments. All electrodes have
a diameter of 18\,mm. They are  aligned by four inner support rods
with ceramic insulators in-between, separating the electrodes by
0.5\,mm axially. The quadrupole rods themselves are mounted via
ceramic supports to four mounting disks as shown in the lower part
of the fig.\,\ref{f_detail_rod}. The inner distance between
opposite quadrupole rods is $2 r_0=12$\,mm. The holding disks are
kept at fixed distances by the four outer support rods. Additional
ceramic rods are provided to ``seal'' the system. This is to
maximize the gas pressure inside the trap while minimizing the
pumping requirements of the whole system. Electrode segments of
various lengths are used (see Table \ref{t_mechanics}). The
shorter segments at the entrance of the trap allow to set up an
axial potential gradient for retardation and focusing of the
incoming ions as discussed above. In the part of the trap where
the potential decreases slowly and linearly, longer segments are
sufficient. A high segmentation is again required in the region
where the ions are finally accumulated and ejected.

\subsection{The extraction section and the pulsed cavity}

\begin{figure}
{\hspace{.2cm}
 \begin{center}
 \epsfig{file=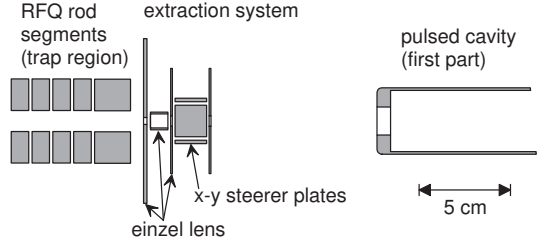,width=7cm}
 \caption{\label{f_extraction_optics}
 Exit side of the linear trap with
 extraction lens, x-y steerers and a pulsed cavity}
 \end{center}
}
\end{figure}

Figure~\ref{f_extraction_optics} shows the extraction optics of
the buncher and the pulsed cavity. The extraction hole has a
diameter of 6 mm. The extraction optics consists of an einzel lens
formed by a cylindrical electrode between two diaphragms  and a
set of x-y steerers. The purpose of this arrangement is to adapt
the divergence of the extracted ion bunch to the ion optics
downstream and to allow adjustments of the ion-flight direction.

At a distance of 90\,mm downstream from the extraction optics, the
ions enter the pulsed cavity, which is only partly shown in the
figure. The cavity is designed to shield the ion pulse from
external fields and has a total length of 380\,mm. This is larger
than the expected length of a pulse of ions with mass number
$A>20$ that is injected into the cavity with an kinetic energy of
2 to 3\,keV. In Fig.\,\ref{f_buncher_setup} it can be seen that
the pulsed cavity is fixed to a support directly mounted on an
insulated flange perpendicular to the beam axis.

\subsection{Vacuum system and gas supply}

\begin{figure}%[b]
{\hspace{.2cm}
 \begin{center}
 \epsfig{file=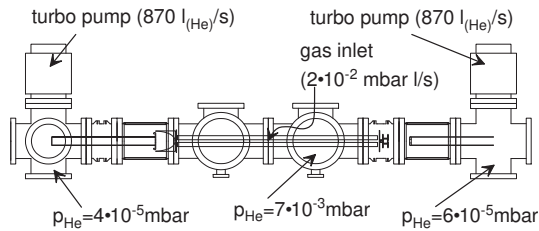,width=7cm}
 \caption{\label{f_vacuum_schematic}
 Schematic of the vacuum system of the ISOLTRAP ion beam cooler
 and buncher. Gas flow value (valve controller reading),
 pumping speeds (for He), and
 measured pressures values (He-corrected) are shown.}
\end{center} }
\end{figure}

Figure\,\ref{f_vacuum_schematic} gives a schematic overview of the
vacuum and gas inlet system. The vacuum system that houses the
ion-optical parts discussed above consists of several vacuum
chambers mostly of CF 150 size. Two magnetic-bearing
turbo-molecular pumps (Pfeiffer TMU 1000MC) with a pumping speed
of 870\,l/s for helium are connected on both sides of the system.
These pumps are backed by one 40\,m$^3$/h roughing pump.
Differential pumping is accomplished by diaphragms at the entrance
and exit side in order to reduce the gas flow out of the trapping
region. The trap is mainly pumped via the entrance and exit holes
for the ions. The buffer-gas inlet is via a needle valve connected
to a pressure controller (Balzers EVR\,116 + RVC\,200). Inside the
vacuum chamber, a thin stainless steel tube  guides the gas into
the central region of the RFQ structure. A full-range pressure
gauge (Balzers PKR\,250) is connected to the vacuum chamber on the
HV platform. Its reading is used as the input for the pressure
controller. The pressure value measured with the gauge is
 of course lower than the pressure inside the structure. From
gas flow simulations, it is expected that the gauge gives a
reading which is about a factor of 10 lower than the actual
pressure inside the center of the structure\footnote{In this
paper, the values given for the measured pressure are the vacuum
gauge readings corrected for the gas-specific factor.}.

A reasonable working pressure inside the RF-structure is
$p_{\mathrm{He}} \approx 10^{-2} $\,mbar. With the resulting gas
load, care has to be taken about possible discharge processes in
the system. A critical point is the region where the ground
injection electrode is close to the deceleration electrode
(Fig.\,\ref{f_retardation_optics}) with a gap of 5 mm  and a
potential difference of up to 60 kV. One has to make sure that at
this gap a pressure regime is reached, where the discharge
probability is minimized. The Paschen curve~\cite{Meek} for helium
gives a maximum pressure of the order of  $10^{-5}$\,mbar. Gas
flow calculations have confirmed that such values can be reached
with the applied pumping speed, the entrance and exit openings of
the linear trap, the desired inside pressure, and the required gas
flow. The experimentally measured pressures are given in
Fig.\,\ref {f_vacuum_schematic}.

Another point of concern are gas impurities in the system which
can give rise to charge exchange and loss of ions. Such effects
have been seen in the start-up of the system when ions like argon
were injected into the ion beam cooler and buncher. Meanwhile,
this has been cured by baking the whole system. Furthermore,
pellets of non-evaporable getter material (SAES St~172 type) have
been installed in the gas inlet line of the buncher for gas
purification.

\subsection{Electronics and control system}
\subsubsection{The high-voltage system}

\begin{figure}
{\hspace{.2cm}
 \begin{center}
 \epsfig{file=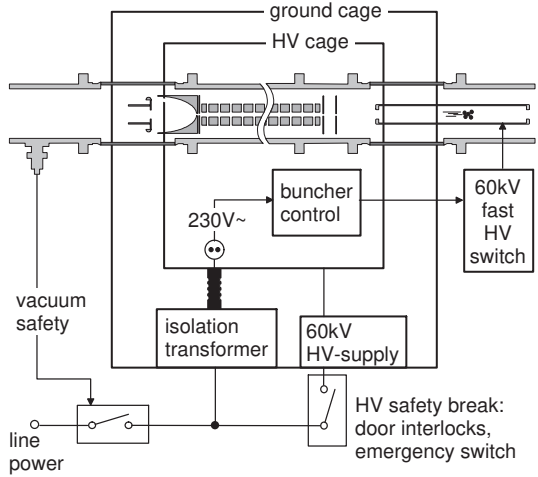,width=7cm}
 \caption{\label{f_HV_power}
 Line power and HV supply for the high-voltage platform. Also shown
 are the safety interlocks that cut the high voltage in the event
 of too high pressure and allow safe access to the cage.}
\end{center}
}
\end{figure}

Figure\,\ref{f_HV_power} shows schematically the high-voltage and
the line power supply for the high-voltage platform. An oil-free
2-kW isolation transformer is used to provide line power to the
platform. The platform potential is provided by a
remote-controllable power supply (FUG HCN 140M-65000) with a
maximum voltage of 65\,kV and a short- and long-term voltage
stability of $10^{-5}$. This power supply is connected to the
secondary windings of the transformer via a 5.6-k$\Omega$
current-limiting resistor. Varistors across the secondary winding
of the isolation transformer protect the electronic equipment on
the high-voltage platform from high-voltage transients caused by
occasional sparkovers.

\begin{figure}%[b]
{\hspace{.2cm}
 \begin{center}
 \epsfig{file=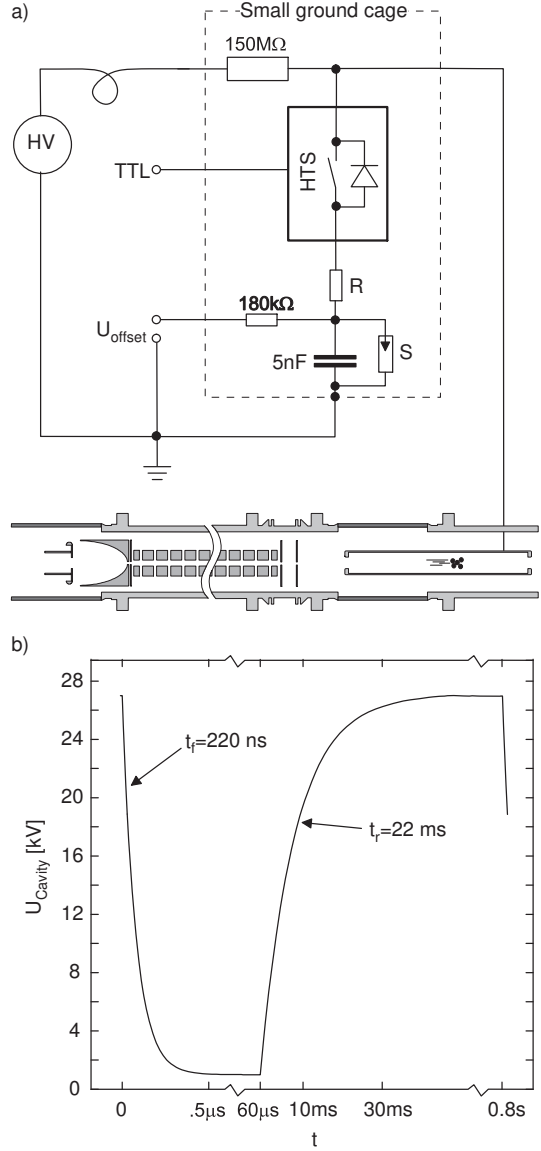,width=7cm}
 \caption{\label{f_HV_drifttube}
 Circuit used for the fast switching of the pulsed cavity from a
 potential of $\leq$60\,kV to ground potential (a). Measured voltage
 for a switching from
 $\approx$30\,keV to ground as a function of time (b).
 HV: high-voltage power supply; HTS: high-voltage transistor switch;
 $U_{\mathrm{offset}}$:
 compensating voltage for the voltage drop over the internal resistance
 of the HTS; S: spark gap.
 The times $t_{\mathrm{f}}$ (fall-time)
 and $t_{\mathrm{r}}$ (rise-time) are the times needed for the signal
 to change between 10 and 90\%
 of its maximum strength.}
 \end{center}
}
\end{figure}

Figure\,\ref{f_HV_drifttube} shows the circuit used for the
switching of the potential of the cavity. A fast high-voltage
transistor switch (Behlke HTS\,650) discharges the cavity with a
fall-time $t_{\mathrm{f}}=220$\,ns. After a constant on-time of
60\,$\mu$s the switch opens again, allowing the cavity to be
charged again to high voltage with a rise-time $t_{\mathrm{r}} =
22$\,ms. Due to the internal resistance of the high voltage
switch, the cavity can only be discharged to ground when a
negative bias voltage of several hundred volts is applied to the
point marked $U_{\mathrm{offset}}$ (a typical value is
$\approx500$\,V). A spark gap S serves to protect the low-voltage
side of the circuit.

\subsubsection{Radiofrequency and DC supplies for the
buncher}

\begin{figure}
{\hspace{.2cm}
 \begin{center}
 \epsfig{file=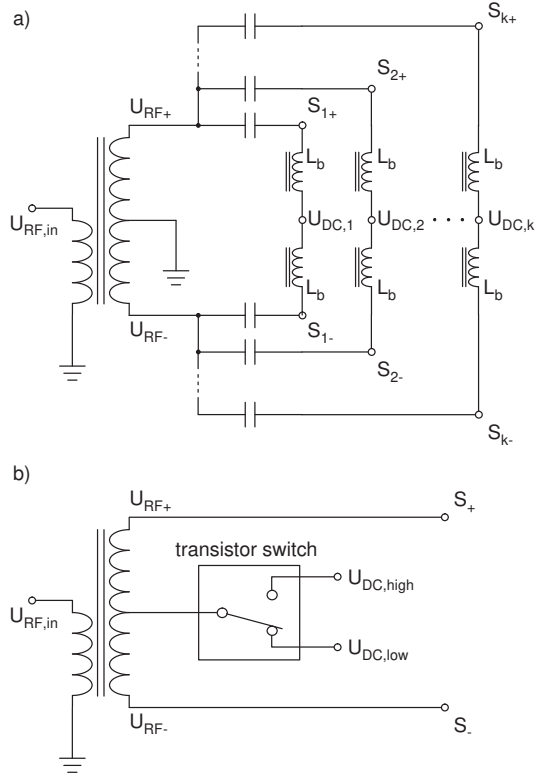, width=7cm}
 \caption{\label{f_RF_DC_circuit}
 Principle of the electronic circuits used to provide the
 different segments $S_{k\, \pm}$ of the
 four rods with DC ($U_{\mathrm{DC},k}$) and
 RF ($U_{\mathrm{RF\,\pm}}$) voltages. a) circuit for
 those electrodes which
 require RF and static DC voltages; b) circuit for those
 electrodes which
 require RF and switched DC voltages.}
\end{center} }
\end{figure}

The RF voltage for the radial pseudopotential has to be applied to
all segments of the system but with a phase shift of 180$^\circ$
between segments of neighboring rods. The DC potentials are
applied in two different modes, depending on whether the element
is used for a static potential or whether the potential is
dynamically switched within less than one microsecond.

Two different types of circuits are used. The circuit shown in
Fig.\,\ref{f_RF_DC_circuit}(a) is used for all elements where a
static DC voltage is required. The RF voltage $U_{\mathrm{RF,in}}$
is provided by a function generator (Stanford Research DS 345)
after amplification by a 200-W radiofrequency amplifier (ENI 240L)
and fed into a transformer. The secondary winding has a grounded
center tap in order to generate two symmetrical radiofrequency
signals $U_{\mathrm{RF} \pm}$ of opposite sign. These RF voltages
are added via capacitors $C=10$\,nF to the DC voltages
$U_{\mathrm{DC},i}$ ($i=1\ldots k$) in order to obtain a combined
RF/DC signal which applied to the different pairs of electrodes
$S_{i\, \pm}$ ($i=1\ldots k$). Two inductances
$L_{\mathrm{b}}=3.9$\,mH protect each DC feeding point from the
two RF signals. The DC voltages are provided by 0 to 500-V power
supply modules (Brandenburg). The voltage series required for the
generation of the linear potential slope (see
Fig.\,\ref{f_buncher_principle}) is obtained via a resistive
voltage divider chain fed by two of the DC power supply modules.
The most critical electrodes closest to the injection side of the
buncher and those in the trapping region have separate voltage
supplies. The inductance of the transformer and the capacitances
have been matched to the capacitance of the electrodes in order to
obtain a maximum RF voltage at a frequency of $\nu \approx$1\,MHz.
An amplitude $U_{\mathrm{RF}}\approx 125$\,V is reached with about
40\,W RF power.

For the extraction of the ions from the buncher it is necessary
to switch the potential of the last segments of the linear trap
very quickly. Therefore, the relevant segments are supplied
with a circuit as shown in Fig.\,\ref{f_RF_DC_circuit}(b). The
DC voltage from the DC modules is sent  to the center tap
of the transformer
via a fast transistor switch. With this
system, a fall-time of 0.5 $\mu$s is obtained.

A number of additional power supplies is used for the voltages of
the deceleration and focusing electrode (see
Fig.\,\ref{f_retardation_optics}) and the extraction optics (see
Fig.\,\ref{f_extraction_optics}). The high voltage
$U_{\mathrm{cavity}}$ for the pulsed cavity (see
Fig.\,\ref{f_HV_drifttube}) is also provided from the HV platform.

\subsubsection{The control system}

\begin{figure}
{\hspace{.2cm}
 \begin{center}
 \epsfig{file=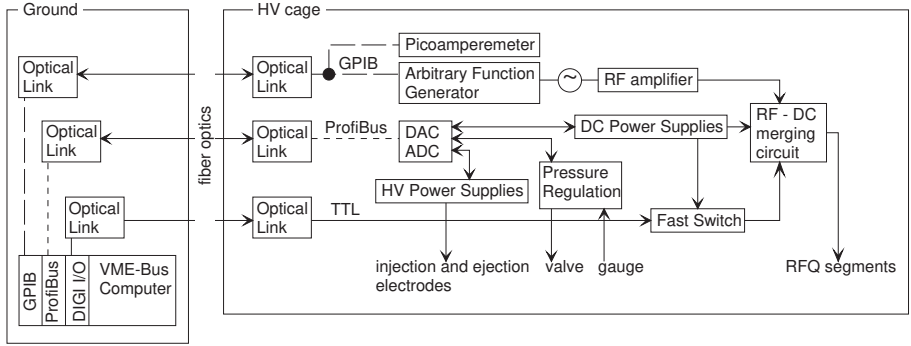,width=12cm}
 \caption{\label{f_control_layout}
 Layout of the control system of the ISOLTRAP buncher. The electronics
 on the HV platform is controlled via three optical links.}
 \end{center}
}
\end{figure}

Figure\,\ref{f_control_layout} shows the layout of those parts of
the control system relevant for the operation of the cooler and
buncher. Since most of the electronics is installed on the HV
platform, remote control of the system is mandatory. Three
different optical links provide the communication between the
VME-based computer system of the ISOLTRAP spectrometer and the
electronics on the HV-platform. A GPIB link (National Instruments
GPIB-140) is used to control the function generator for the
radiofrequency signal and to read out a pico-amperemeter. A field
bus system (Profibus) equipped with DACs and ADCs is used for
analog programming and monitoring a total of 28 voltage supplies.
It also serves for programming the pressure regulation system. A
fast TTL link (Harting) is used to provide the trigger for the
fast voltage switching required for the ion extraction from the
buncher.

\subsubsection{Diagnostics}

Several tools are available for the optimization and diagnosis of
the performance of the ion beam buncher. The ISOLDE beam scanner
and movable Faraday-cup in front of the buncher allow the
measurement of profile and current of beams from both ISOLDE and
the test ion source. In order to optimize the injection of the
60-keV ion beam into the buncher, it is important to perform a
beam current measurement on the high-voltage platform. For this
purpose a pico-amperemeter (Keithley 485) is installed on the
platform which can be connected to any of the buncher electrodes.
On the ejection side another Faraday cup and a micro-channel plate
(MCP) detector are available. This detector consists of 2 MCP's
with 18 mm active area mounted in a Chevron arrangement. No
electron repelling grid is used (see discussion in section
\ref{kap_efficiency}). This MCP detector has been used for most of
the test measurements reported below. More MCP detectors are
available further downstream for the tuning of the ion beam
transport to the Penning trap system. In addition, a beam viewing
system is available which can be installed behind the 90$^\circ$
vertical deflector (see fig.\,\ref{f_isoltrap}) to obtain the
spatial distribution of the ion beam. This system (Colutron
BVS-2), which is mounted on a flange with a viewport, consists of
an MCP with a phosphor screen behind it. The fluorescent light
from the phosphor screen originated by the secondary electrons of
the MCP after ion impact, can be viewed through this window with a
CCD camera. When this detector is used the deflector electrode
system is removed from its vacuum chamber in order to allow the
ions to pass through.

\section{Operation}

Table \ref{t_buncher_parameters} gives a typical set of operating
parameters for ions with mass number close to $A=39$ and $A=133$.
The ions enter the linear trap with an energy of 20\,eV after
retardation. The value given for the buffer gas pressure value is
the (helium-corrected) vacuum gauge reading. As discussed above
the pressure inside the linear trap is estimated to be a factor of
10 higher. Depending on the delivered ion beam current, the ions
are accumulated for a period $T_{\mathrm{accu}}= 0.01 \ldots
1000$\,ms. Then the ions are allowed to complete their cooling
into the trap potential minimum for an additional period
$T_{\mathrm{cool}}= 2 $\,ms before they are ejected. The timing of
the switching of the cavity is adapted to the mass-dependent time
of flight of the ions. For a 2.65-keV extraction and an $A=40$
ion, $t_{\mathrm{switch}}= 12\,\mu $s. The system has been tested
with repetition rates up to 5 Hz. In general, repetition rates of
$\approx$1\,Hz are used for the mass measurements.

\begin{table}
\caption{\label{t_buncher_parameters} Typical operating parameters
of the cooler and buncher for ions in the mass ranges $A\approx
39$ and $A\approx 133$.}
\begin{center}
\vspace*{1cm}
\tiny
\scriptsize
\begin{tabular}{llrl}
\hline \multicolumn{2}{l}{Parameter}  &
\multicolumn{2}{l}{Value}\\ \hline \hline
\multicolumn{2}{l}{helium pressure (at gauge position)
$p_{\mathrm{He}}$ } &$6\cdot 10^{-3}$&mbar\\ \multicolumn{2}{l}{RF
frequency $\nu_{\mathrm{RF}}=\omega_{\mathrm{RF}}/2\pi$} & 970
&kHz\\ RF amplitude  $U_{\mathrm{RF}}$         &    for
$A\approx133$ & 135    &V\\ &    for $A\approx39$       &  97 &V\\
cooling time $T_{\mathrm{cool}}$        &    for $A\approx133$ &
10     &ms\\ &    for $A\approx39$           &  2      &ms\\
\multicolumn{2}{l}{cage voltage $U_{\mathrm{HV}}$}          &  30
&kV\\[3mm] \multicolumn{3}{l}{electrode voltages
$U_{\mathrm{elec}}$ relative to the cage voltage
$U_{\mathrm{HV}}$:}\\[2mm] \multicolumn{2}{l}{deceleration
electrode}         &  --1350  &V\\ \multicolumn{2}{l}{focusing
electrode}             &  --230   &V\\[2mm] quadrupole rod segment
&{\# 1 }                    &  --60      &V\\ &{\# 2}&  --40 &V\\
&{\# 3}                              &  --25     &V\\ &{\# 4 to \#
22}                 &  --10  to --20  &V\\ &{\# 23 (accumulation)}
&  --24     &V\\ &{\# 23 (ejection)}              &   +2
&V\\
      &{\# 24, \# 25 }                 &  --29     &V\\
      &{\# 26 (accumulation)}      &    0          &V\\
      &{\# 26 (ejection)}              &  --55         &V\\[2mm]
\multicolumn{2}{l}{plates of the extraction system} &  --420
&V\\ \multicolumn{2}{l}{einzel lens }                         &
--55    &V\\ \multicolumn{2}{l}{pulsed cavity
($U_{\mathrm{offset}} =$\,--500\,V,  see Fig. 15)}
                                   &  --2650 &  V\\
\hline
\end{tabular} \tiny
\normalsize
\end{center}
\end{table}

\section{Performance of the ISOLTRAP ion beam buncher}

A number of systematic investigations have been performed with the
buncher in order to characterize its performance and to compare
the results with simulations. For these measurements, both ions
from ISOLDE and from the test ion source have been used.

\subsection{Injection, capture, and ejection}

\subsubsection{Ion energy}
A critical parameter for the injection and capture of the ions
into the buncher is the energy of the ions when they enter the
system. This energy is determined by the potential of the linear
trap on the HV platform with respect to the initial ion beam
energy. If the potential is too high, then the ions are not able
to enter the system.  If it is too low, then the energy loss in
the gas is not sufficient to prevent the ions from leaving the
system on the exit side or to be stopped on the extraction
diaphragm.

\begin{figure}[tb]
{\hspace{.2cm}
 \begin{center}
 \epsfig{file=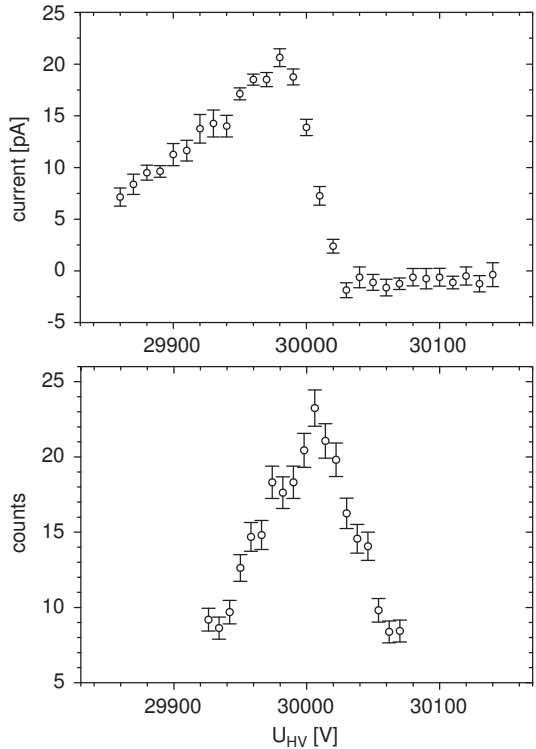,width=7cm}
 \caption{\label{f_HV_potentialscan}
 Current of $^{132}$Xe$^+$ ions guided
 through the RFQ (top) and number of  accumulated and ejected ions in
 bunching mode (bottom)
 as a function of the  voltage $U_{\mathrm{HV}}$ of the HV platform.}
 \end{center}
}
\end{figure}

In order to optimize the injection and the platform potential, two
kinds of measurements are performed. In the first case, all
electrodes of the extraction system (see Fig.
\ref{f_extraction_optics}) are connected together for a direct
current measurement. Radiofrequency and DC potentials are applied
to the segmented rods as given in Tab.\,\ref{t_buncher_parameters}
for the ejection mode. No buffer-gas is used.

In Fig.\,\ref{f_HV_potentialscan} (top) the result of such a
measurement is shown for a 30-keV $^{132}$Xe$^+$-ion beam as a
function of the potential of the HV platform. At the ion beam
energy of 30\,keV a maximum in the transmission is observed.
Increasing the potential from 30\,000\,V to 30\,030\,V gives a
sharp drop in transmission. At a potential larger than 30\,030\,V
no ions are transmitted. For potentials lower than 30\,kV and
correspondingly higher ion beam energy in the system, the
transmission decreases gradually. This can be understood since the
focusing of the ion optics in the injection part was optimized for
a beam entering the cooler and buncher at about 20\,eV kinetic
energy.

This transmission mode of operation is particularly useful for a
fast tuning of the beam injection into the cooler and buncher. The
platform potential is set to the value giving maximum transmission
and then the current at the end of the structure is monitored as a
function of focusing and steering voltages applied to ion optical
elements in the ISOLDE beam line.

For fine tuning of the injection (in particular of the platform
potential), the full accumulation-ejection cycle has to be
employed. The result of such a measurement (performed with a
buffer gas pressure of $p_{\mathrm{He}}=1\cdot 10^{-2}$\,mbar) is
shown in the bottom part of Fig.\,\ref{f_HV_potentialscan}.
Plotted as a function of the platform voltage are the average
number of $^{132}$Xe$^+$ ions per cycle extracted from the buncher
and detected by the MCP detector. It should be noted that the
optimum value for the platform potential has shifted by about
25\,eV, which means that a maximum capture efficiency requires an
injection of the ions at an energy slightly lower than that for
best transmission.

\subsubsection{Radiofrequency parameters}
\begin{figure}
{\hspace{.2cm}
 \begin{center}
 \epsfig{file=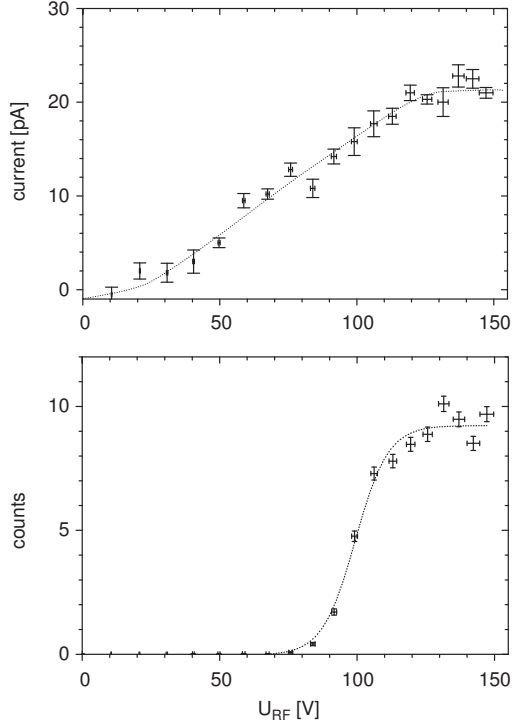,width=7cm}
 \caption{
 \label{f_ampscan}
 Transmission of $^{132}$Xe$^+$ through the
 cooler and buncher (top) and number of  accumulated and ejected
 ions (bottom) as a
 function of the radiofrequency amplitude $U_{\mathrm{RF}}$. The
 lines shown serve to guide the eye. }
\end{center} }
\end{figure}

Both, the amplitude and the frequency of the RF potential should
have an effect on the ion transmission through the linear trap and
on the storage of the ions. Due to technical reasons, the
frequency of the RF circuit used in this work was fixed to
$\nu_{\mathrm{RF}}\approx$1\,MHz. Therefore, only measurements on
the effect of a change of the RF amplitude were performed. The
result of such a measurement is shown in Fig.\,\ref{f_ampscan}.
The upper part shows an injection and transmission measurement
similar to those described above. The ion current at the end of
the structure is plotted as a function of the amplitude
$U_{\mathrm{RF}}$ of the applied radiofrequency. As expected, the
transmission steadily increases with increasing amplitude (and
deeper radial pseudopotential). Above $U_{\mathrm{RF}}\approx
120$\,V, the transmitted current saturates. The lower figure shows
the result of a measurement where the Xe ions are accumulated,
cooled ($p_{\mathrm{He}}=1\cdot 10^{-2}$\,mbar) and ejected. Shown
is the number of detected ions as a function of $U_{\mathrm{RF}}$.
It can be seen that a certain threshold amplitude of
$U_{\mathrm{RF}}\approx 75$\,V is required in order to obtain ions
from the trap. The reason is that for too low RF amplitudes, the
defocusing radial part of the DC potential at the axial potential
minimum of the trap is stronger than the generated
pseudopotential, as explained in section 3.1. The ions are
transmitted to the ``trapping'' region where they are finally lost
radially. Using (\ref{U_critical}) and the actual trap parameters,
one obtains a threshold amplitude of $U_{\mathrm{RF,min}} \approx
60$\,V which is in good agreement with the experimentally observed
value.

\subsubsection{Buffer gas pressure}

\begin{figure}
{\hspace{.2cm}
 \begin{center}
 \epsfig{file=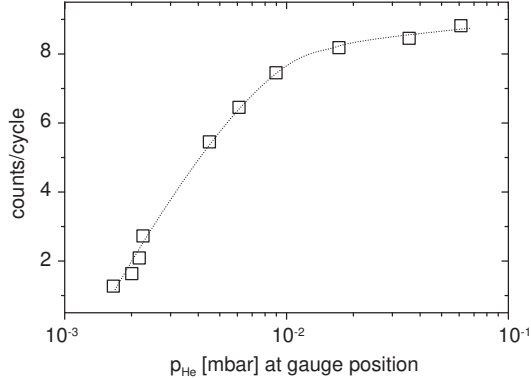,width=7cm}
 \caption{\label{f_pressscan}
 Number of accumulated and ejected $^{208}$Pb$^+$ ions  as a
 function of the buffer gas pressure $p_{\mathrm{He}}$. The
 value is the
 (He-corrected) reading of the gauge at the vacuum chamber
 containing the linear ion trap. The line shown serves
 to guide the eye.}
 \end{center}
}
\end{figure}

The accumulation of the ions in the linear ion trap requires
sufficient energy dissipation in the buffer gas (see
Fig.\,\ref{f_cooling_principle}) and hence a high enough pressure.
If the pressure is too low, then most of the ions will hit the
exit electrode of the linear ion trap and are lost.
Figure\,\ref{f_pressscan} shows the number of ions ejected from
the trap for constant injection conditions as a function of the
gas pressure. It can be seen that a He pressure of a few
$10^{-3}$\,mbar is required in order to observe any ions. Above
$p_{\mathrm{He}}\approx 10^{-2}$\,mbar the number that can be
extracted starts to saturate.

\subsection{\label{kap_efficiency}Efficiency}

After finding good parameters for the operation of the cooler and
buncher the efficiency of the system was determined. From a
current measurement as shown in Fig.\,\ref{f_HV_potentialscan}
(top) and a comparison with the beam current measured with the
Faraday-cup in front of the system, its injection and guiding
efficiency $\epsilon_{\mathrm{trans}}$ was determined. Such
measurements have been carried out with ISOLDE beams from various
ion sources.  Values for the transmission
$\epsilon_{\mathrm{trans}}^{\mathrm{exp}}$ between 20\% and 40\%
were found. This is in good agreement with the theoretical value
of $\epsilon_{\mathrm{trans}}^{\mathrm{theo}}\approx 35$\,\% (see
Section \ref{sec_decelinj}).

%It should be emphasized that the measured value includes the
%transport of the ions over nearly the full length of the buncher.

Most interesting is the total efficiency
$\epsilon_{\mathrm{total}}$ of the cooler and buncher, which is
defined as the ratio of the number of ions injected into the
system and the number of ions finally counted in the extracted ion
pulse. The measurements are performed in the following way: an
attenuated ISOLDE beam, but still intense enough to perform a
reliable current measurement (typically a few pA), is transported
to the apparatus. For the injection into the buncher, a very short
opening time for the ISOLDE beam gate is used in order to inject
only a well defined small number of ions $N_{\mathrm{ion}}$ into
the buncher. These ions are accumulated, cooled, ejected, and
finally detected with the MCP detector. As mentioned above, this
detector, which is normally only used for simple monitoring
purposes, is not equipped with an electron-repelling grid.
Therefore, according to \cite{Brehm95}, only about half of the
maximum possible efficiency of $\approx 50$\, \% is achieved in
the case of moderately heavy ions with 2.5 keV energy. With the
number of counted events being $N_{\mathrm{MCP}}$ and an
efficiency $\epsilon_{\mathrm{MCP}}\approx 30\,\%$, the overall
efficiency is given by
$\epsilon_{\mathrm{total}}=
 (N_{\mathrm{MCP}}/\epsilon_{\mathrm{MCP}})/N_{\mathrm{ion}}$.

Total efficiency measurements have been performed for various Xe
isotopes and values of $\epsilon_{\mathrm{total}}\approx
12\ldots15$\,\%  were achieved. This value allows to study ions
from the weakest radioactive beams available at ISOLDE with the
ISOLTRAP spectrometer. However, compared to the efficiency
$\epsilon_{\mathrm{trans}}$ for injection and transmission, the
total efficiency $\epsilon_{\mathrm{total}}$ is lower by a factor
of two to three. Storage time measurements for alkali ions have
shown lifetimes of up to several 100 ms. For Xe ions as used for
the efficiency measurement, no significant charge exchange with
gas impurities has been observed, which could have caused ion
loss. A possible explanation is that the MCP efficiency is lower
than assumed. It is also likely that an optimal match of ISOLDE
beam emittance and buncher acceptance has not yet been achieved.
This will be investigated in forthcoming test measurements
accompanied by further beam injection simulations.

\subsection{Cooling and beam emittance}

\subsubsection{Damping of ion motion and cooling}

\begin{figure}
{\hspace{.2cm}
 \begin{center}
 \epsfig{file=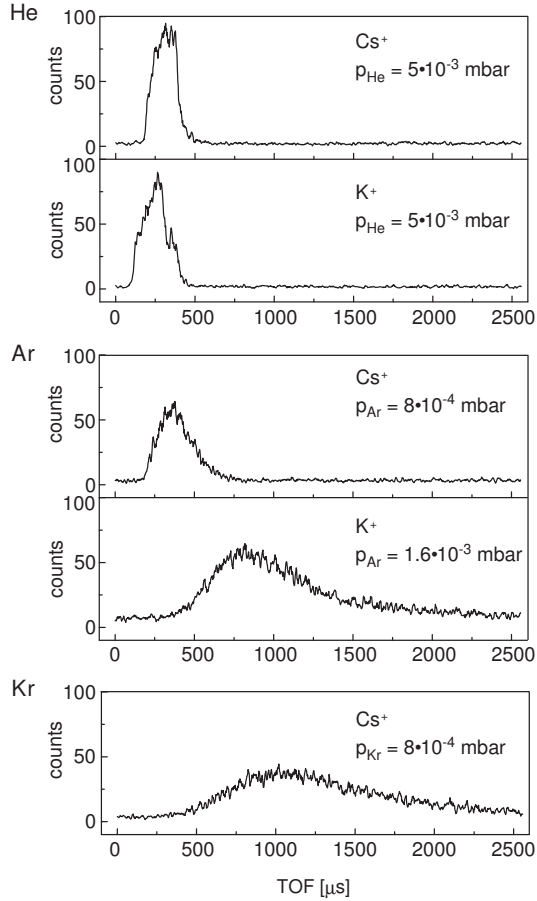,width=7cm}
 \caption{\label{f_ionpuls_damping}
 Time structure of pulses of different ions after passage
 through the RFQ system for
 different buffer gases. Here, the ion trap was operated as an
 ion guide.}
 \end{center}
}
\end{figure}

In order to illustrate the damping of the axial ion motion in the
cooler and buncher for different ions and types of buffer gas,
very short ion pulses have been injected. The DC potential of the
cooler was permanently set to the extraction mode, which means
that the ions were directly accelerated to the MCP detector after
a single passage through the system.

The result of such a measurement for various ion species and gases
is shown in Fig.\,\ref{f_ionpuls_damping}. With buffer gas of
higher mass, the ion pulses arrive considerably later and are
broadened. This can be understood by looking at the microscopic
beam simulations shown in Fig.\,\ref{f_visc_micro_comparison}.
There the trajectories show a significant scattering also for the
axial motion. In the experiment the transmission of $^{39}$K$^+$
ions through $^{84}$Kr was also tried, but, as expected no ion was
able to reach the detector.

\begin{figure}
{\hspace{.2cm}
\begin{center}
\epsfig{file=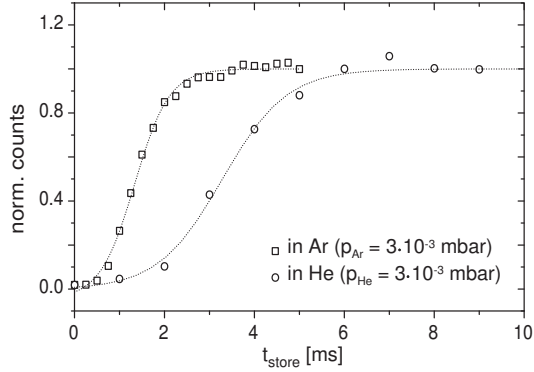,width=7cm}
\caption{\label{f_countscool} Normalized number of  $^{133}$Cs$^+$
ions extracted from the trap as a function of the storage time
inside the trap. The measurements were performed with helium and
argon as buffer gas. }
\end{center} }
\end{figure}

An illustration of the transverse cooling process as a function of
time is given in Fig.\,\ref{f_countscool}. In the experiment,
$^{133}$Cs$^+$ ions were injected into the linear trap for a short
period  $T_{\mathrm{accu}}=2 \mu$s. The figure shows the number of
ions ejected out of the trap as a function of the cooling time
$T_{\mathrm{cool}}$ after injection. This measurement was
performed for helium and argon as buffer gases at about the same
pressure. For short cooling times, the number of ions that can be
extracted is small and only above $T_{\mathrm{cool}}> 300 \mu$s is
the maximum number reached. This is because at the pressure of
$p_{\mathrm{He}}\approx 3\cdot 10^{-3}$\,mbar used here several
oscillations are required inside the trap  before the radial
extent of the ion cloud is smaller than the radius of the
extraction hole. The figure also illustrates that stronger
damping, and as a consequence faster cooling, is provided by the
heavier buffer gas.

\begin{figure}
{\hspace{.2cm}
\begin{center}
\epsfig{file=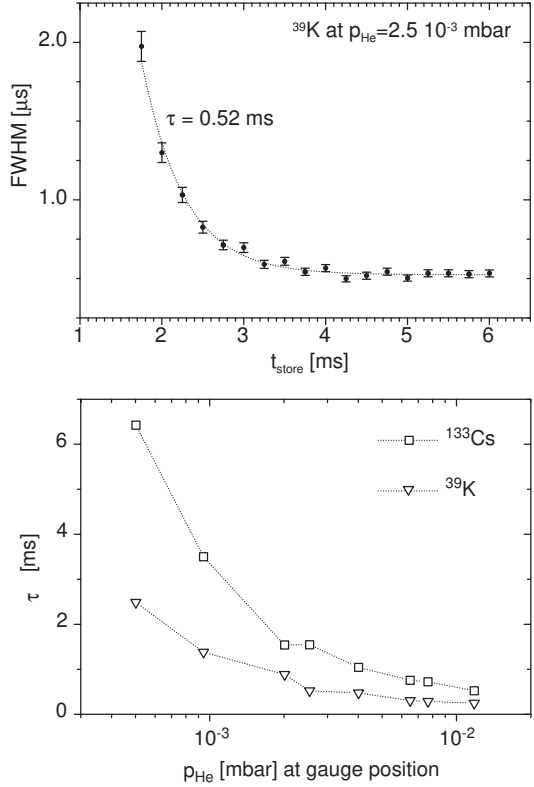,width=7cm} \caption{\label{f_fwhmcool}
Top: temporal width of ejected $^{39}$K ion pulses as a function
of the storage time inside the trap. An exponential is fitted to
the data. Bottom: time constants obtained as shown in the top
figure as a function of the helium buffer gas pressure for K and
Cs ions. }
\end{center} }
\end{figure}

Figure\,\ref{f_fwhmcool} illustrates once more  the axial cooling
process. In the top figure the temporal width of the ejected ion
pulse is shown as a function of the cooling time for the case of
$^{39}$K$^+$ ions and $p_{\mathrm{He}}= 2.5 \cdot 10^{-3}$\,mbar.
As can be seen in the figure, an exponential decrease of the pulse
width with time is observed. The pulse width decreases with
decreasing ion temperature and smaller axial distribution of the
ions in the trap. The width is therefore a good indicator of the
ion temperature. The time constant $\tau_{\mathrm{cool}}$ of the
curve can to first order be regarded as a measure of the cooling
time constant. The value of $\tau_{\mathrm{cool}}=0.5$\,ms
obtained in the measurement shown in Fig.\,\ref{f_fwhmcool} is
comparable to a value of 0.3 ms calculated from ion mobility data
and an estimate of the pressure inside the buncher close to the
trapping region of $p\approx 10^{-2}$\,mbar.
Figure\,\ref{f_fwhmcool} also shows that the cooling time of the
ions in the trap should be several times $\tau_{\mathrm{cool}}$ in
order to extract ion pulses with minimum temperature out of the
trap. The bottom part of Fig.\,\ref{f_fwhmcool} shows the measured
time constants $\tau_{\mathrm{cool}}$ for Cs and K ions as a
function of the pressure. For the highest pressures, time
constants down to a few hundred microseconds are observed.

\subsubsection{Cooling limit and emittance}

\begin{figure}
{\hspace{.2cm}
 \begin{center}
 \epsfig{file=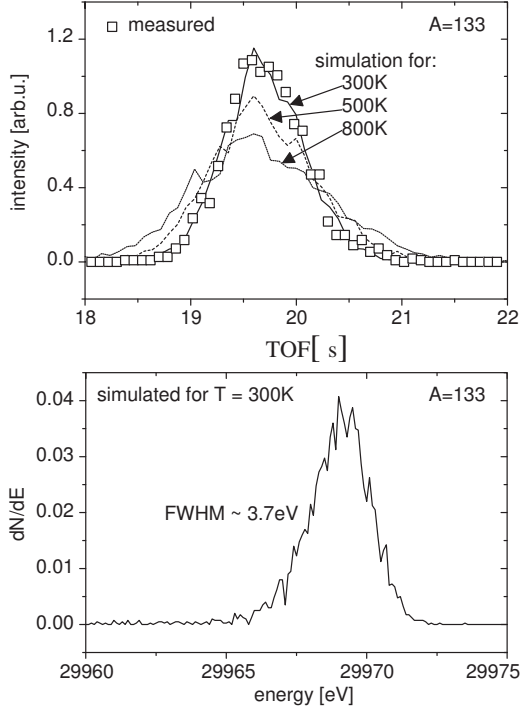,width=7cm}
 \caption{\label{f_ionpuls_temperature}
 Top: temporally resolved 30-keV ion pulse (squares)
 of $^{133}$Cs$^+$ ions compared
 to theoretical ion pulse shapes corresponding to initial ion
 temperatures of 300\,K, 500\,K and 800\,K. The theoretical curves are
 normalized to the same number of ions. Bottom: calculated energy
 spectrum of the ion pulse corresponding to the 300-K pulse
 shape shown in the top figure.}
\end{center} }
\end{figure}

For a determination of the final ion temperature both in the axial
and radial direction, two kind of measurements have been carried
out. In both cases, ions were accumulated and cooled for
$T_{\mathrm{cool}}= 20$\,ms to reach their equilibrium
temperature.

For the determination of the axial temperature, the number of ions
in a pulse has been measured temporally resolved with the MCP
detector. The measured shape of the ion pulse can be compared to
calculated shapes based on ion distributions at different
temperatures. Such pulse shapes are shown in the top part of
Fig.\,\ref{f_ionpuls_temperature} for $^{133}$Cs ions extracted
from the linear trap and accelerated to 30\,keV. The data are
plotted with respect to the time of ejection from the trap. The
points correspond to the experimental data, the curves to
calculated shapes for ion distributions with temperatures of
300\,K, 500\,K, and 800\,K. A good agreement is achieved for a
temperature of 300\,K giving evidence that the ions reach the
temperature of the buffer gas. This is in accordance with the
predictions of Fig.\,\ref{f_ion_distributions}. The bottom part of
Fig.\,\,\ref{f_ionpuls_temperature} shows the calculated energy
distribution of the ejected ions. For the calculation, the
confirmed  300-K ion distribution was taken together with the
electric fields for the extraction as used in the experiment. From
the time and energy spread of the ion pulses shown in the figure,
a value for the longitudinal emittance of the ion pulse ejected
from the buncher can be obtained. The total value is
$\epsilon_{\mathrm{long}}\approx 10$\,eV\,$\cdot \mu \rm s$.

The transverse temperature of the ions can be extracted from a
beam emittance measurement. Such a measurement has been performed
in the following way. The beam observation system was mounted at
the 90$^\circ$ deflector chamber (Fig. \ref{f_isoltrap} with the
deflector removed). Ion pulses with an energy of 2.5 keV were
created by using the pulsed cavity. With an einzel lens in front
of the deflector chamber, the beam was focused onto the detector
and the beam profile was observed with a CCD camera. The profile
was found to be Gaussian-shaped with 90\,\% of the beam within a
radius of about 1.8 mm. A beam scraper mounted on a linear
feed-through at a distance of 500 mm in front of the beam viewing
system was used to determine the size of the beam at this
position. There a beam radius of 2.5 mm (corresponding to
$>90$\,\% beam intensity) was observed.  From this a beam
divergence of 5 mrad is derived. Combining both measurements gives
a rough upper limit for the transverse beam emittance of
$\epsilon_{\mathrm{trans}}\approx 10 \pi$\,mm\,mrad.

If the extracted pulse was accelerated to 60 keV instead of 2.5
keV, it would have an emittance of about $2\pi$\,mm\,mrad, which
corresponds to more than tenfold improvement with respect to the
original ISOLDE beam. This shows that gas-filled radiofrequency
ion guides and traps can very effectively be used to improve the
emittance of ion beams.

\section{Future developments}

\begin{figure}
{\hspace{.2cm}
 \begin{center}
 \epsfig{file=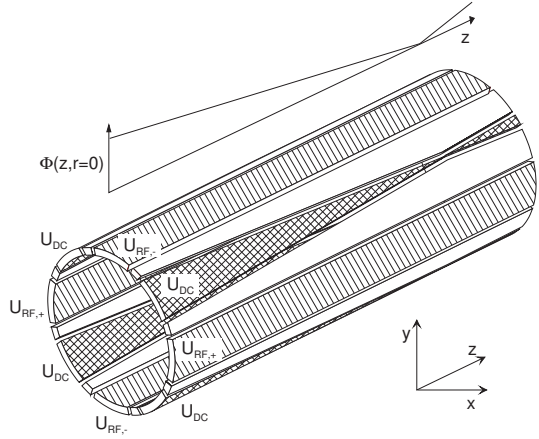,width=7cm}
 \caption{\label{f_cylbunch}
 Principle layout of a cylindrical radiofrequency
 quadrupole ion trap. Straight electrodes (line-hatched)
 are used to create the RF quadrupole field.
 An axial trapping potential is produced via tapered electrodes
 (cross-hatched) and corresponding ground electrodes (white).
 $\Phi(z,r=0)$ sketches the expected potential on the axis
 of the cylinder.}
\end{center} }
\end{figure}

A new type of radiofrequency quadrupole ion beam cooler and
buncher has been investigated recently. It will be briefly
discussed here while a comprehensive study including simulations
will be presented in \cite{Schwarz2000misc}.

Figure\,\ref{f_cylbunch} shows the principal layout of the system,
which is based on a segmented cylinder. Four segments of constant
width are used to create the radiofrequency quadrupole field by
applying RF voltages $U_{\mathrm{RF}}$ with appropriate phases. An
axial trapping potential $\Phi(z)$ is produced by applying
DC-voltages $U_{\mathrm{DC}}$ to eight tapered electrodes, which
are  placed between grounded counter electrodes. For the ejection
of accumulated ions, the voltage applied to the four short tapered
electrodes is switched to a lower value.

The advantage of this system is, compared to the present system,
that the RF and DC voltage supplies are decoupled and no matching
circuit like the one shown in Fig.\,\ref{f_RF_DC_circuit} is
needed. Furthermore, only four electrical feedthroughs are
required for the system shown in Fig.\,\ref{f_cylbunch}. However,
higher RF and DC voltages have to be applied, which will require a
very careful design of the electrodes.

Information gained from the present setup, from additional ion
optical simulations, and from voltage breakdown tests with
prototype electrode systems will be used as input for a detailed
design of such a new simplified system of the ISOLTRAP ion beam
cooler and buncher.

\section{Conclusions and Outlook}
In the work presented here, we have demonstrated for the first
time the accumulation, bunching, cooling and emittance improvement
of radioactive ion beams from an on-line mass separator by means
of a linear radiofrequency ion trap. The efficiency of the system
and the properties of the energy-variable ion bunches are in
agreement with the theoretical expectations.

Compared to the previously used Paul trap system
\cite{SchwarzPhD}, the efficiency of ISOLTRAP has been increased
by three orders of magnitude. This enables ISOLTRAP to be applied
to very exotic isotopes with low production rates. First
measurements of this kind have already been successfully
performed. The investigation of the masses of neutron-deficient
mercury isotopes was extended to $A=182$~\cite{Schwarz2000misc}. A
mass measurement of the very short-lived ($T_{1/2}=173$\,ms)
isotope $^{33}$Ar became possible \cite{Herfurth2000}, delivered
by ISOLDE with an intensity of only a few thousand atoms per
second. With this increased sensitivity, a very fruitful mass
measurement program will be possible in the future.

The work has again confirmed that the concept of using trapping
techniques to manipulate radioactive ion beams is very successful.
Based on the experience with the present linear ion trap system we
study now the design of a larger beam buncher for the preparation
of short low-emittance ion pulses with close to 100\,\%
efficiency. With a pulsed cavity as used in this work, the energy
of the pulses can easily be varied from nearly zero up to more
than 100 keV. The variability of this energy is expected to be
very useful for nuclear-decay study experiments, laser
spectroscopy experiments on unstable isotopes, as well as for
solid-state and surface physics studies using radioactive probes.

Furthermore, the use of linear radiofrequency quadrupole ion
accumulators, coolers and bunchers will play an important role in
schemes presently discussed for the stopping of relativistic beams
of nuclear reaction products and their conversion into low-energy
ion beams.

%Two recent examples of an ongoing or and a planned project
%are the SHIPTRAP project at GSI/Darmstadt, where low energy beams of
%heavy and superheavy elements will be produced or JYFLTRAP at
%Jyv\"askyl"a, or the
%RIA radioactive ion beam project presently under discussion in the US.

\begin{ack}
The authors wish to thank J.~Bernard, W.~Hornung, G.~Marx and
W.~Quint at GSI Darmstadt, F. Ames and P. Schmidt from 
the university of Mainz and the CERN summer students S.~Lindner,
S.~Harto, and C.~Richter for their valuable help during the
development and commissioning of the system. This work was partly
carried out within the EXOTRAPS project in the EU LSF-RTD program
and financially supported under contract number ERBFMGECT980099.
It was also supported by NSERC of Canada.
\end{ack}

%\bibliographystyle{amsnotitsh}
%\bibliography{isoltrap,pauletc,linpaul,isolde,mansetc,gases,misc}
%\providecommand{\bysame}{\leavevmode\hbox to3em{\hrulefill}\thinspace}

\end{document}